\documentclass[]{article}  
\usepackage{spie}  
\usepackage{psfig}  

\def\la{\hbox{\rlap{\raise.3ex\hbox{$<$}}\lower.8ex\hbox{$\sim$}\ }}
\def\ga{\hbox{\rlap{\raise.3ex\hbox{$>$}}\lower.8ex\hbox{$\sim$}\ }}
\def\cts{cts~cm$^{-2}$~s$^{-1}$~keV$^{-1}$}

\title{Balloon Flight Background Measurement with Actively-Shielded 
Planar and Imaging CZT Detectors} 

\author{P. F. Bloser\supit{a}, T. Narita\supit{b}, J. A. Jenkins\supit{b}, 
M. Perrin\supit{c}, R. Murray\supit{b},
and J. E. Grindlay\supit{b}
\skiplinehalf
\supit{a}Max-Planck-Institut f\"ur extraterrestrische 
Physik, Giessenbachstrasse, D-85748 Garching, Germany
\skiplinehalf
\supit{b}Harvard-Smithsonian Center for Astrophysics, 60 Garden St., 
Cambridge, MA 02138, USA
\skiplinehalf
\supit{c}Astronomy Department, University of California at Berkeley, 
Berkeley, CA 94720, USA
}

\authorinfo{Further author information: (Send correspondence to
P. Bloser)\\P.B.: E-mail: bloser@mpe.mpg.de}

\begin{document} 
\maketitle 

\begin{abstract}
We present results from the flight of two prototype CZT detectors on a
scientific balloon payload in September 2000.   The first detector,
referred to as ``CZT1,'' consisted of a 10 mm $\times$ 10 mm $\times$ 2 mm 
CZT crystal
with a single gold planar electrode readout.  This detector was
shielded by a combination of a passive collimator in the front, giving
a 40 degree field  of view and surrounded by plastic scintillator, and
a thick BGO crystal in  the rear.   The second detector, ``CZT2,'' comprised two
10 mm $\times$ 10 mm $\times$ 5 mm CZT crystals, one made of eV Products high
pressure Bridgman material and the other of IMARAD horizontal Bridgman
material, each fashioned with a 4 $\times$ 4 array of gold pixels on a 2.5 mm
pitch.  The pixellated detectors were flip-chip-mounted side by side and read out
by a  32-channel ASIC.  This detector was also shielded by a
passive/plastic collimator in the front, but used only additional
passive/plastic shielding in the rear.  Both experiments were flown from
Ft. Sumner, NM on September 19, 2000 on a 24 hour balloon flight.  
Both instruments performed well.  CZT1 
recorded a non-vetoed background level at 100 keV of $\sim 1 \times 10^{-3}$ \cts.
Raising the BGO
threshold from 50 keV to $\sim 1$ MeV produced only an 18\% increase in this
level.  CZT2 recorded
a background at 100 keV of $\sim 4 \times 10^{-3}$ \cts 
~in the eV Products detector and $\sim 6 \times 10^{-3}$ \cts ~in the IMARAD
detector, a difference possibly due to our internal background subtracting procedure.
Both CZT1 and CZT2 spectra were in basic agreement with Monte Carlo simulations,
though both recorded systematically higher count rates at high energy than predicted.
No lines were observed, indicating that neutron capture 
reactions, at least those producing decay lines at a few 100 keV, are not 
significant components of the CZT background.
Comparison of the CZT1 and CZT2 spectra indicates that passive/plastic
shielding may provide adequately low background levels for many applications.
\end{abstract}



\section{INTRODUCTION}
\label{sec:intro}  

Hard X-ray and gamma-ray detectors for high energy astronomy 
made of Cadmium-Zinc-Telluride (CZT) are finally realizing
the potential that they have shown for the past
6--7 years.  It has long been known that CZT detectors offer 
far better energy resolution than scintillator detectors such as 
NaI and CsI, and that the use of pixel or strip electrode readouts allows far
better spatial resolution.  Response up to 600 keV is possible with 
moderate thicknesses (5 mm) due to the high stopping power of CZT, and
no cryogenic cooling is required due to its wide band gap.  Extensive
laboratory work by our group at Harvard\cite{bloser98mrs,narita98,narita99,narita2000}
and many others has now begun to bring this promise to fruition.  In the
hard X-ray band ($\sim 10$ to several 100 keV) the first
space-based astronomical instrument employing CZT, the Swift 
mission\cite{barthelmy}, is already under construction. 
EXIST\cite{grindlay2000}, a wide-field, all-sky survey telescope operating 
between $\sim 5$ and 600 keV, is in the early planning stages, and the 
Hard X-ray Telescope of
Constellation-X\cite{constellationx}, a narrow-field instrument using
focusing optics up to 40 keV, is under development.  CZT is
also being considered for the medium gamma-ray band (0.5--50 MeV) as the 
calorimeter material in advanced Compton telescopes\cite{tigre}.

A major contributor to the maturity of CZT detector technology has been
an active campaign of balloon flight tests over the past five 
years, both by our group\cite{bloser98,bloser99,bloser2000} and by 
others\cite{parsons96,harrison98,slavis98,slavis99,slavis2000}.  These 
experiments are necessary to test detector
design, readout, and shielding methods under flight conditions, and to 
measure background levels in the space environment.  The experiments have
progressed from simple one-element detectors with passive 
shielding\cite{parsons96,harrison98} to single-element detectors with
various combinations of active and passive 
shielding\cite{parsons96,bloser98} to complex multi-anode detectors, either
strips\cite{slavis98,slavis99,slavis2000} or pixels\cite{bloser99,bloser2000},
again using both passive and active shielding techniques.  The general 
conclusions that may be drawn from this experience are that CZT detector
technology is indeed compatible with the needs of balloon and space-flight
systems, and that background levels can be made acceptable for sensitive
astronomical observations.  More specifically, active shielding has been 
shown to reduce background significantly over passive shielding 
alone\cite{parsons96,slavis98,slavis99}.  This suggests a substantial
portion of the CZT background is generated by processes involving charged
particles and/or prompt gamma reactions.  It has been suggested that 
prompt (n,$\gamma$) reactions in CZT could be an important source of 
background due to the large neutron capture cross sections in 
Cd\cite{parsons96,harrison98}.  
Additionally, activation of Cd and Te by neutrons could produce isomeric
states that decay radiatively on a timescale too long to be vetoed by active
shields\cite{harrison98}.  However, flight results with a strip detector from 
Washington University, St. Louis and the University of California, San
Diego (WUSTL/UCSD) saw no evidence
of decay lines\cite{slavis98}.  In addition, the WUSTL/UCSD experiment
found that a shield threshold of 10 MeV, too high to veto prompt gamma-rays,
still reduced background to acceptable levels\cite{slavis99}.  Perhaps, then,
only charged particle shields made out of plastic scintillator are necessary,
instead of heavier gamma-ray shields such as CsI or BGO.

In this paper we present results from two CZT experiments
conducted in September 2000: the first array of 
pixellated CZT detectors, read out by an ASIC, to be flown on a balloon,
and a single-element background experiment flown
simultaneously.  Motivated by the
requirements of the EXIST concept\cite{grindlay2000}, our goals were to
investigate techniques for assembling a large-area detector plane out of
small elements, test various detector materials in the near-space 
environment, demonstrate an ASIC readout system in flight, and compare
active shielding techniques using particle and photon shields for a 
wide field-of-view telescope.  

\section{Description of the CZT Experiments}
\label{sec:description}

\begin{figure}
\begin{center}
\begin{tabular}{lr}
\psfig{figure=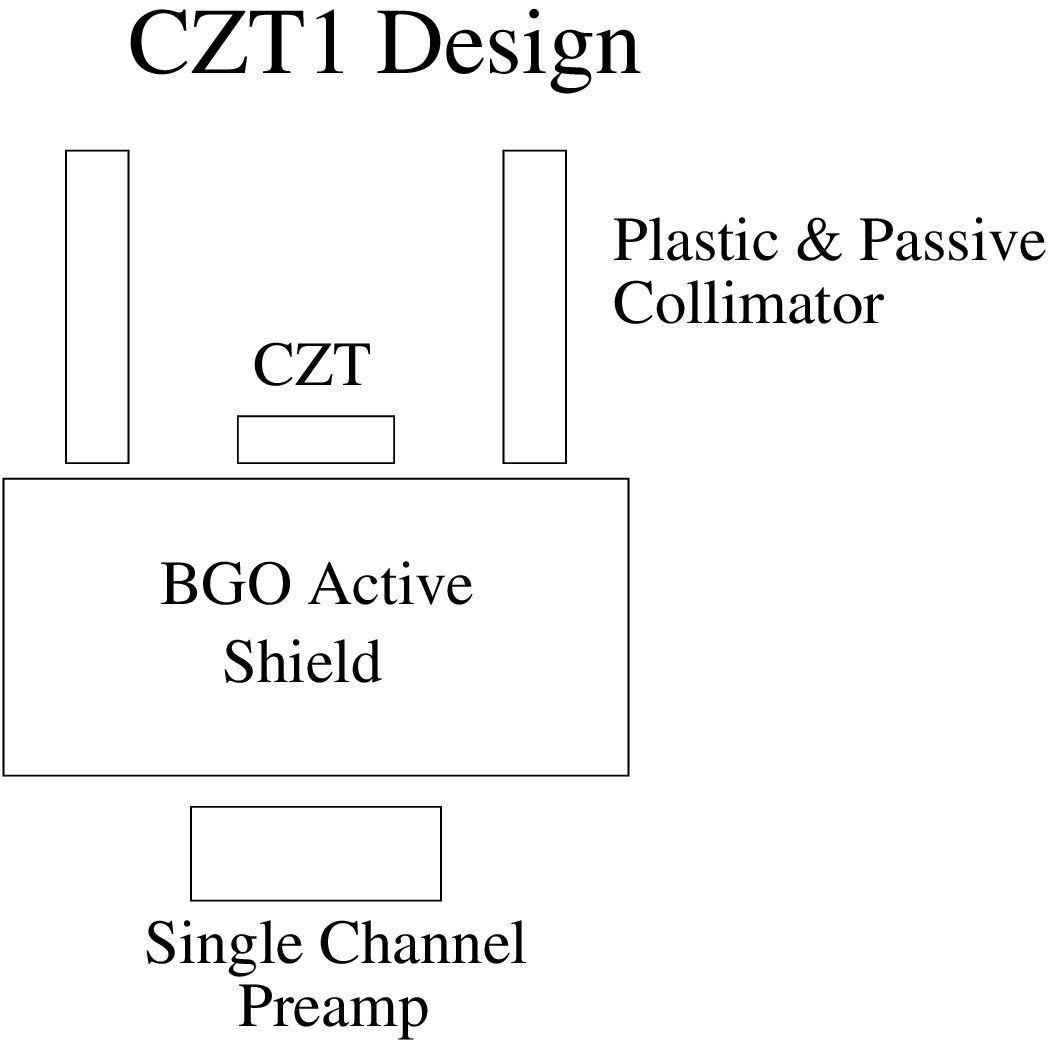,height=2.1in,width=3.2in}
& \psfig{figure=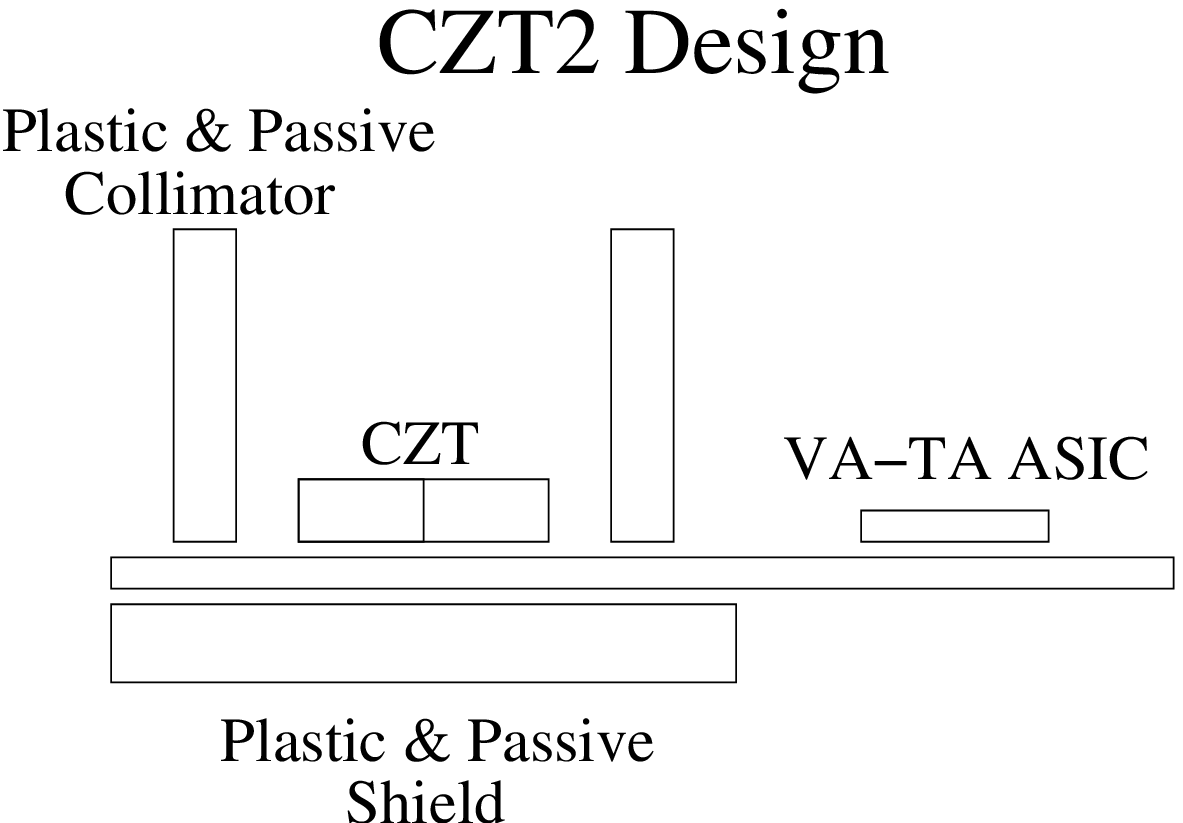,height=2.1in,width=3.2in}
\end{tabular}
\end{center}
\caption[initial]
{\label{fig:schematics}
Schematic diagrams of the CZT1 ({\em left}) and CZT2 ({\em right}) detectors
flown in September 2000.  CZT1 is a single-element detector with a 
passive/plastic collimator and a rear BGO active photon shield.  CZT2 consists of
two pixellated detectors read out by an ASIC in an identical passive/plastic
collimator, but with only additional passive/plastic shielding to the rear.
}
\end{figure}

The motivations for our two experiments and the detector designs
have been described in detail previously\cite{bloser99,bloser2000}.  Here
we repeat the main features, as shown schematically 
in Figure~\ref{fig:schematics}.

The first CZT experiment, dubbed ``CZT1,'' is a modification of the CZT/BGO detector
flown by us in 1997\cite{bloser98}.  Its main purpose was to test the effectiveness
of photon vs. particle-only shielding in a wide field of view (FOV) instrument.  
CZT1 consists of a single-element detector,
10 mm $\times$ 10 mm $\times$ 2 mm, supplied
by eV Products and made from high-pressure Bridgman (HPB) material with planar gold
contacts.  The detector sits directly in front of a thick BGO crystal, 8.2 cm
diameter $\times$ 6.5 cm thick, which acts as an active shield.  The threshold of
the BGO was alternated between $\sim 50$ keV and $\sim 1$ MeV to simulate photon
and particle rejection, respectively.
The CZT is surrounded
by a graded Pb-Sb-Cu collimator (4.5 mm Pb, 1 mm Sn, 1 mm Cu) that provides 
a $40^{\circ}$ FOV, appropriate
to an all-sky survey telescope such as EXIST.  This passive collimator is in turn 
surrounded by 0.5'' thick NE-102 plastic scintillator, read out by miniature PMTs,
to reject local gamma production.  Only one
readout channel was available in addition to the CZT channel, and so the signals from
the BGO and the passive/plastic collimator were combined into a single veto.

The second experiment, ``CZT2,'' was intended to test technological requirements
for the EXIST concept and to provide a particle-only shielding configuration to
compare with CZT1.  CZT2 consists of a tiled ``array'' of two flip-chip-mounted pixellated 
detectors, each 10 mm $\times$ 10 mm $\times$ 5 mm.  One detector is made of HPB CZT
provided by eV Products, and the other is horizontal Bridgman (HB) material produced
by IMARAD Imaging Systems\cite{cheuvart90}.  Our work with IMARAD HB CZT has shown that
is is more uniform than HPB material, and that simple gold contacts greatly reduce
leakage current to levels similar to HPB values\cite{narita99,narita2000}.  
Both detectors were fashioned with a
4 $\times$ 4 array of gold pixels on a 2.5 mm pitch plus a guard ring; the layout was
such that the pitch is preserved across the two tiled detectors.  The detectors are
mounted side-by-side in a flip-chip fashion on a ceramic carrier board and read out
by a self-triggering, 32-channel VA-TA ASIC supplied by IDE Corp.  A passive/plastic
collimator identical to that in CZT1 provides a $40^{\circ}$ FOV.  Unlike CZT1, 
however, CZT2 is shielded from the rear by a simple passive/plastic shield.  A weak
$^{241}$Am source placed just above the collimator supplied in-flight calibration
at 60 keV.  

\section{Balloon Flight Results}
\label{sec:results}

Both the CZT1 and CZT2 experiments were launched on September 19, 2000 from
Ft. Sumner, NM as part of a multi-experiment balloon payload that also included
Harvard's EXITE2 hard X-ray telescope and MSFC's hard X-ray
focusing optics experiment HERO.  The flight was quite successful, achieving
24 hours at float altitudes between 115,000 ft and 128,000 ft.
The CZT1 detector was intermittently noisy, possibly due to pickup in the plastic
shield PMT preamps, and care was taken to extract periods
of clean operation as judged by the total count rate.
CZT2 performed well throughout the flight.

\subsection{CZT1 Results}
\label{sec:czt1res}

\begin{figure}
\begin{center}
\begin{tabular}{c}
\psfig{figure=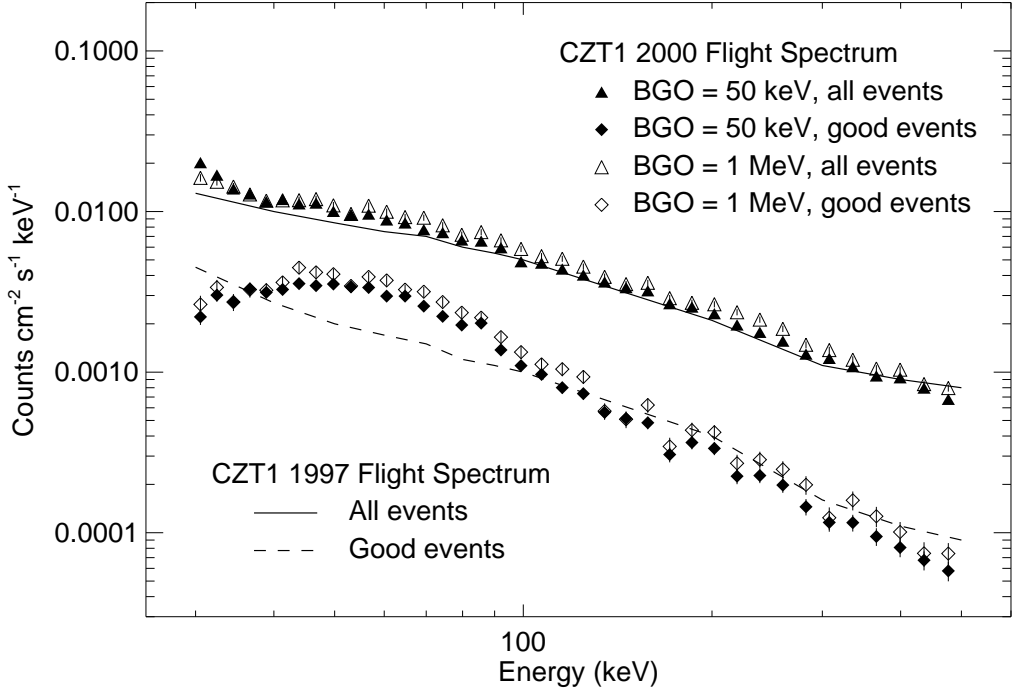,height=10cm}
\end{tabular}
\end{center}
\caption[initial]
{\label{fig:czt1data}
Background spectrum recorded by the CZT1 experiment during the September 2000 flight.
Shown are both total and ``good'' events for BGO threshold settings of 50 keV (closed
symbols) and 1 MeV (open symbols).  The background level at 100 keV for the lower
BGO threshold setting is $\sim 1 \times 10^{-3}$ \cts.
At the higher BGO threshold the background at
100 keV increases by $\sim 18$\%.  Also plotted are the total and good background
levels recorded by the 1997 CZT/BGO experiment\cite{bloser98}.  
}
\end{figure}

The September 2000 flight spectrum recorded by the CZT1 experiment is shown 
in Figure~\ref{fig:czt1data}.
Plotted are both the total and ``good,'' or non-vetoed, events.  Data are included 
from the entire flight, as no altitude-dependent variations within the range of float
altitudes was found previously\cite{bloser98}.  Plotted separately are 5.8 hours of 
data with the BGO threshold set at $\sim 50$ keV (closed symbols) and 3.3 hours of
data with the BGO threshold set at $\sim 1$ MeV (open symbols).  Also shown for
reference are the background levels recorded by the same detector as part of the 
CZT/BGO experiment flown in 1997\cite{bloser98}.  

It is immediately clear that the 2000 background is in very good agreement with the
background recorded in 1997 except at energies below $\sim 100$ keV, where the larger
aperture flux seen in the $40^{\circ}$ FOV is apparent.  In 1997 the CZT detector was
completely covered by a graded Pb-Sn-Cu cup to approximate the grammage of a collimator.
At 100 keV the measured good event level in the 2000 data, for a BGO threshold of 50 keV, 
is $\sim 1 \times 10^{-3}$ \cts.  The BGO provides a factor of $\sim 4.5$ reduction at
this energy, again showing that active shielding is essential.  What is also 
noteworthy is that the background level at 100 keV increases by only $\sim 18$\% when 
the BGO threshold is increased to 1 MeV.  The increase
is approximately constant at all energies.  This confirms the previously-reported result
that active photon shields may be replaced with lighter plastic charged particle 
shields with only a small penalty in background rate\cite{slavis99}.  

\subsection{CZT2 Results}
\label{sec:czt2res}

\subsubsection{In-flight Detector Performance}
\label{sec:czt2performance}

\begin{figure}
\begin{center}
\begin{tabular}{lr}
\psfig{figure=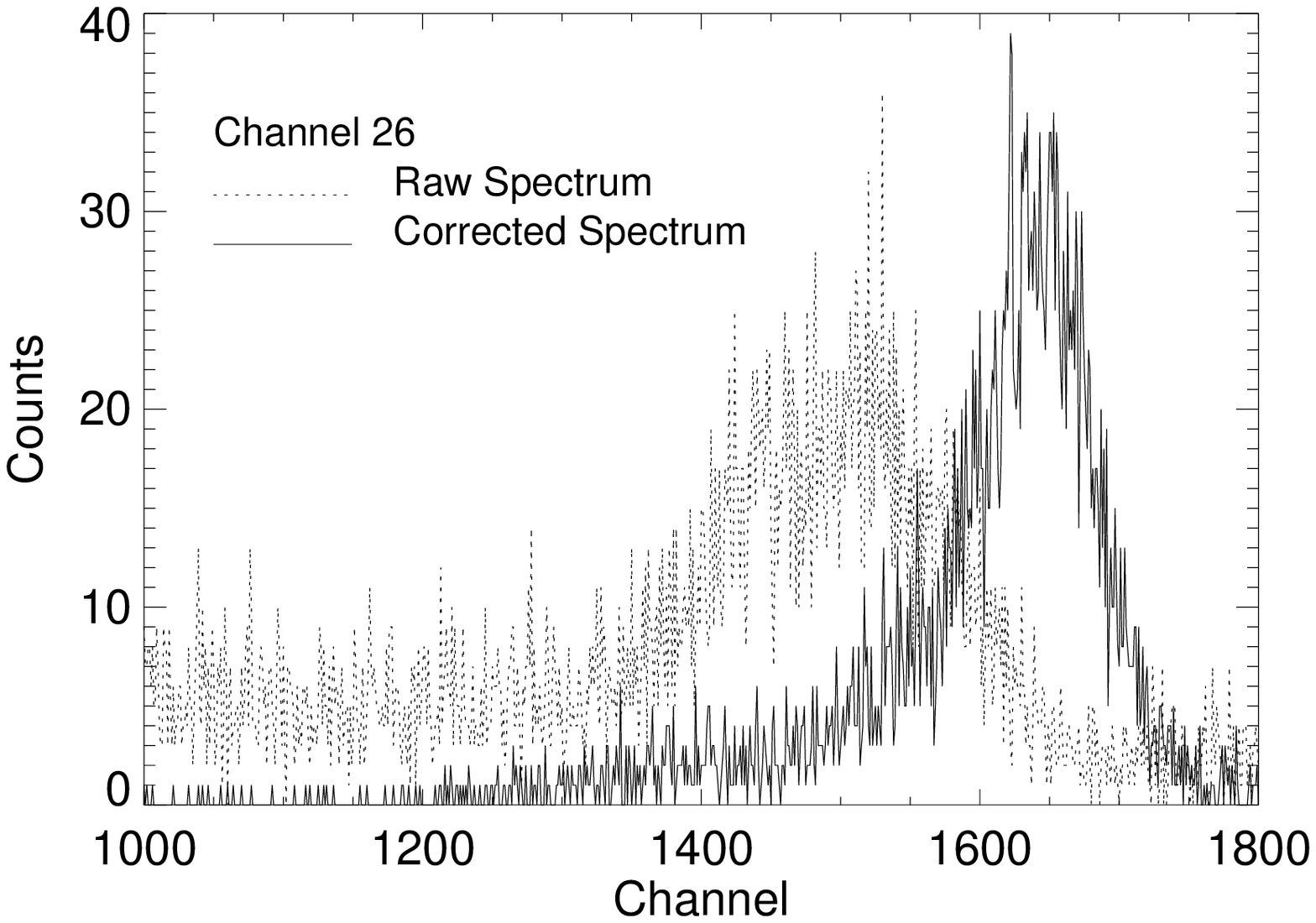,height=10cm,width=8.3cm}
& \psfig{figure=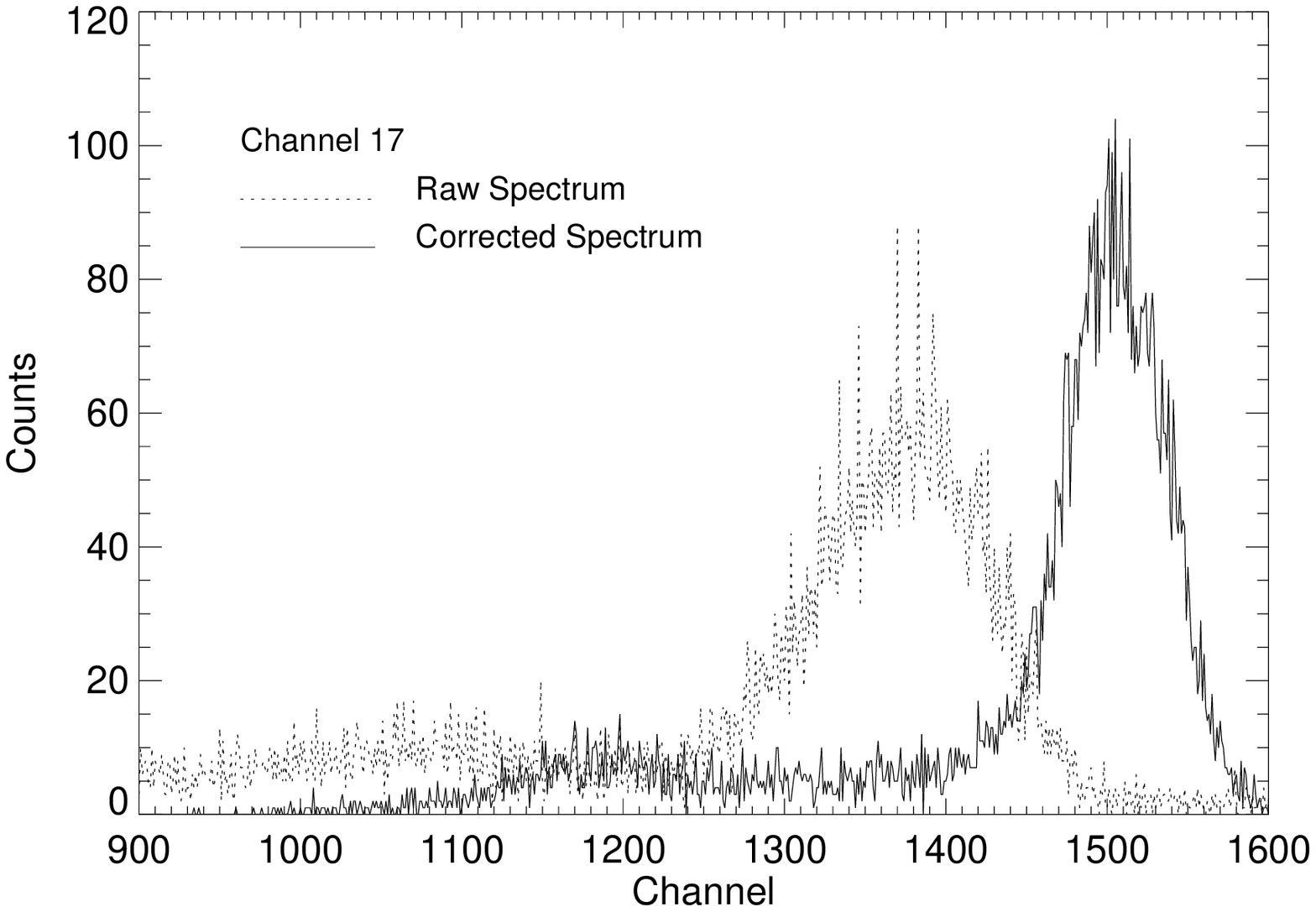,height=10cm,width=8.3cm}
\end{tabular}
\end{center}
\caption[initial]
{\label{fig:performance}
In-flight spectra of the $^{241}$Am 60 keV calibration source, showing the improvement
achieved by adding charge from nearest-neighbor pixels.  At left is a pixel from
the IMARAD HB CZT detector, with a corrected energy resolution of 13.3\% and a photopeak
efficiency of 85\%.  At right is a pixel from the eV Products HPB CZT detector,
with an energy resolution of 9.4\% and a photopeak efficiency of 92\%.  
While the photopeak efficiencies are consistent with those measured on the ground,
the energy resolutions are slightly worse.
}
\end{figure}

One of the primary goals for CZT2 was to demonstrate the in-flight performance of two
pixellated CZT detectors read out by an ASIC.  Although more than half of the channels 
of the IMARAD detector were unusable due to bad contacts after flip-chip mounting or
noise from long lead lengths\cite{bloser2000}, a few of the IMARAD HB CZT pixels 
and all of the eV Products HPB CZT pixels functioned properly.  (The functioning
IMARAD pixels, unfortunately, happened to correspond to areas of the detector with 
poor material qualities\cite{bloser2000}.)
The performance may
be judged by the spectra recorded from the $^{241}$Am calibration source.  An example
from each detector is presented in Figure~\ref{fig:performance}.
Both spectra were corrected for the effects of charge spreading by adding in charge
from the nearest-neighbor pixel, as described previously\cite{bloser2000}.  The 
improvement in the spectra is clear.  The spectra were fit with a combination of 
a gaussian photopeak and exponential tail to determine energy resolution and photopeak 
efficiency, where the photopeak efficiency is defined as the ratio of photopeak counts
to total counts within a range around the gaussian center from $-4\sigma$ to 
$+2.35\sigma$\cite{narita98}.  
At left is an IMARAD pixel with an energy
resolution (corrected spectrum) of 13.3\% at 60 keV and a photopeak efficiency
of 85\%.  At right is an eV Products pixel with an energy resolution of 9.4\% and
a photopeak efficiency of 92\%.  While the photopeak efficiencies are consistent
with those measured on the ground, the energy resolutions are slightly worse; 
at 60 keV we measured 12.9\% for the IMARAD pixel and 8.5\% for the eV Products pixel 
in pre-flight calibrations.  The degradation is small, however.  We conclude that 
a pixellated detector with ASIC readout may be
made to function properly under spaceflight conditions.

\subsubsection{Temperature Effects}
\label{sec:temperature}

\begin{figure}
\begin{center}
\begin{tabular}{lr}
\psfig{figure=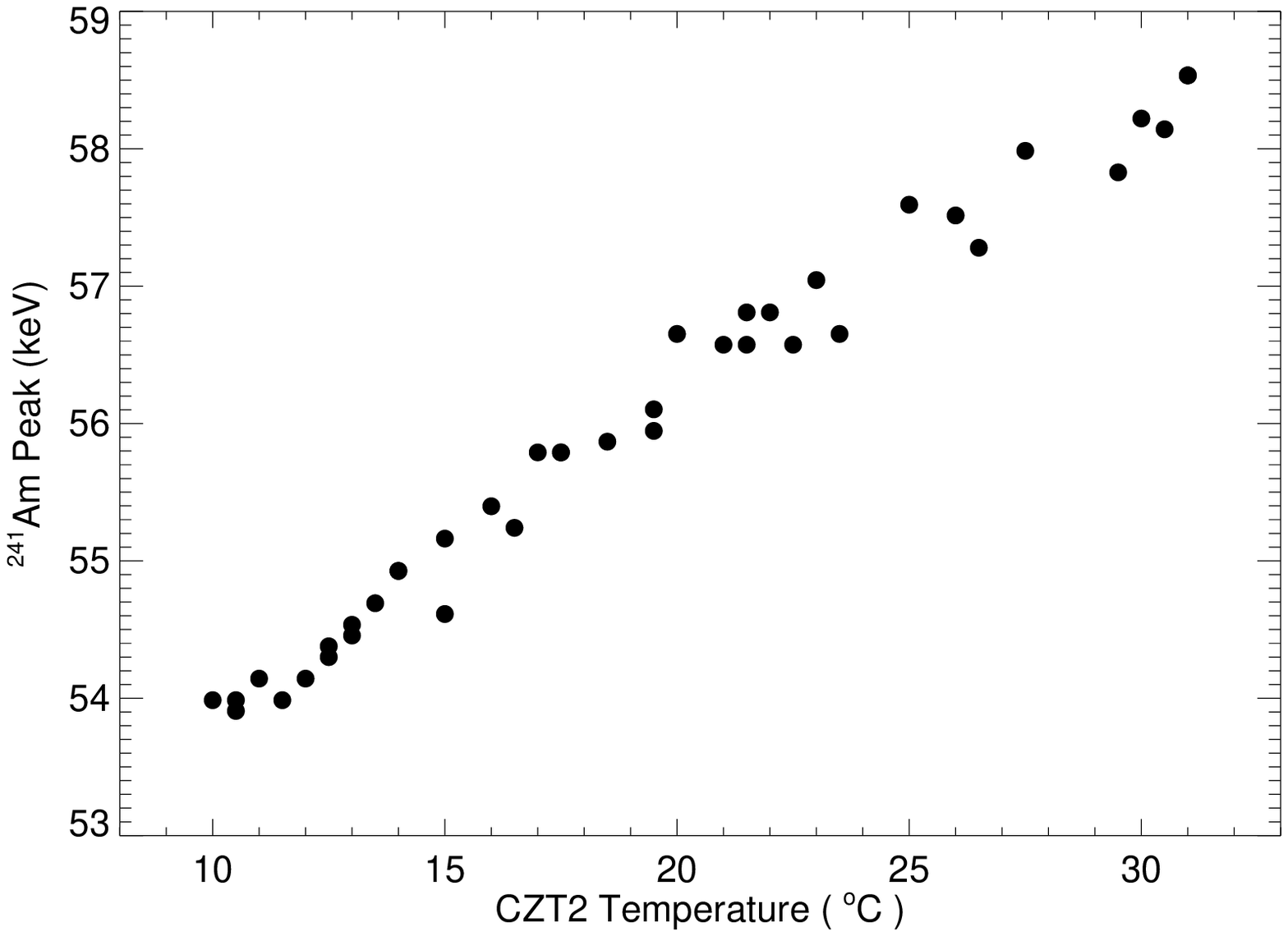,height=8cm,width=8.3cm}
& \psfig{figure=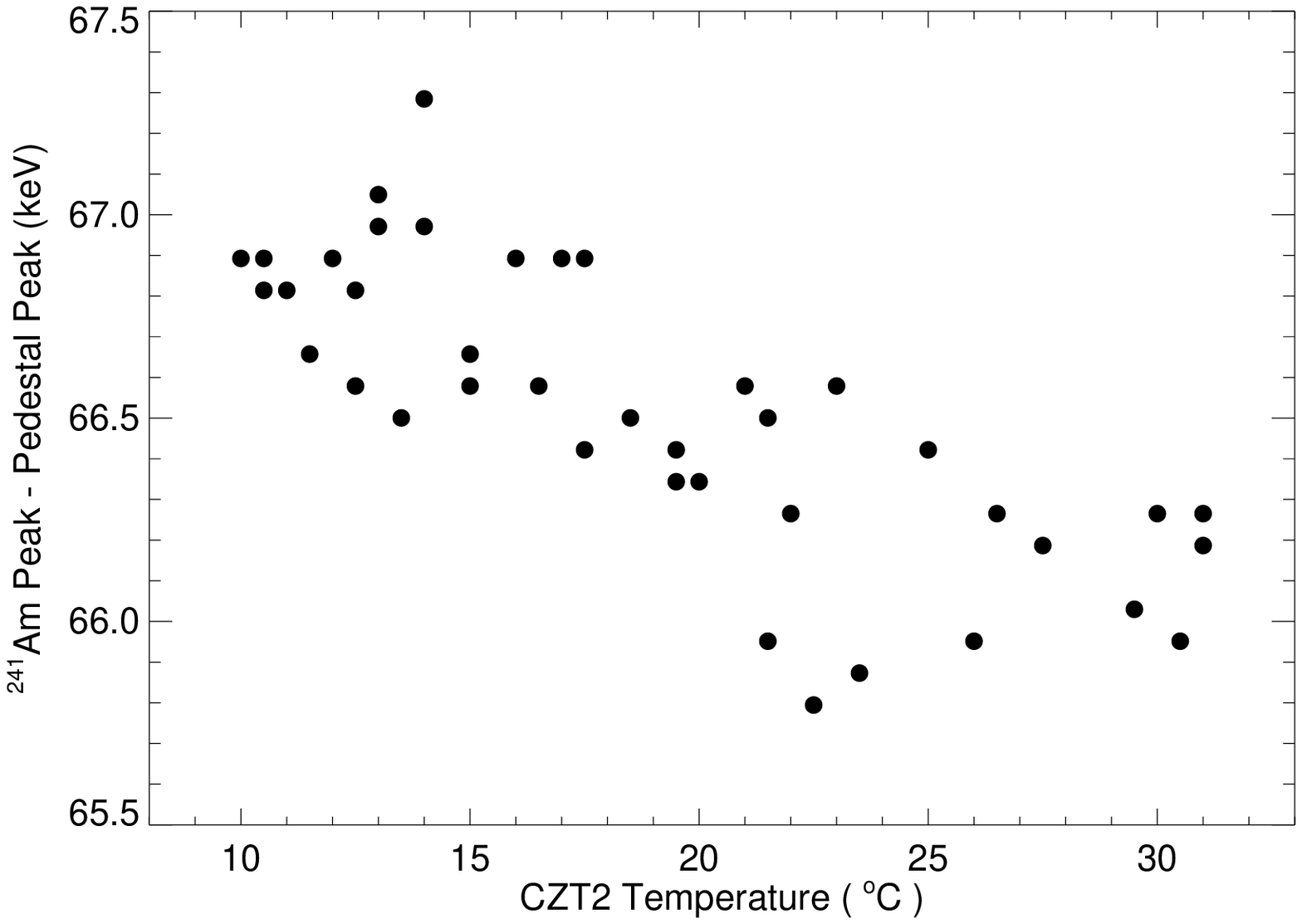,height=8cm,width=8.3cm}
\end{tabular}
\end{center}
\caption[initial]
{\label{fig:temp}
Scatter plots showing the relation between the $^{241}$Am line peak and
the temperature ({\em left}), and the difference between the $^{241}$Am peak
and the pedestal peak and the temperature ({\em right}).  The line peaks are
expressed in energy using the pre-flight calibration.
There is clearly a very
strong correlation between the temperature and the channel of the 60 keV peak.  The
difference between the $^{241}$Am peak and the pedestal peak changes by a much smaller
amount, however, indicating that it is the offset, rather than the gain, that is 
strongly dependent on the temperature.
}
\end{figure}

A temperature sensor was included in the CZT2 experiment to monitor the sensitivity of
the detector gain, offset, and noise to temperature.  During the course of the flight the
temperature inside the CZT2 pressure vessel varied between 10 $^{\circ}$C and 
31 $^{\circ}$C.  
The CZT2 data were assembled into 30 minute segments, and the peak channel and 
width $\sigma$ of both the $^{241}$Am line and the offset pedestal peak were determined 
for an arbitrary channel for each segment.  Scatter plots of 
these quantities revealed no correlation between the width $\sigma$ of the line or 
pedestal and
the temperature.  There is, however, a strong correlation with temperature for the 
peak channel of both
the $^{241}$Am line and the pedestal.  In
Figure~\ref{fig:temp} we show scatter plots of the $^{241}$Am peak vs. 
temperature (left) and the difference between the $^{241}$Am peak and the pedestal peak
vs. temperature (right).  
The line peaks are expressed in keV using the pre-flight energy calibration.  It is 
clear that the calibration has changed in flight, and that this change is highly
correlated with the temperature; the $^{241}$Am line moves between 54 keV and
58.5 keV within this temperature range.  However, the difference between the line
peak and the pedestal peak, though also correlated with temperature, changes far
less, only by about 1 keV.  (Note that according to our pre-flight energy calibrations, 
which 
made use of three different line energies, the pedestal peak is not located exactly
at 0 keV.)  This indicates that it is the offset, rather than the gain, that is most
affected by the temperature.  This means that, to first order, we may correct the
energy of the in-flight spectra by a simple shift in energy offset.  It does indicate,
however, that some degree of temperature control or electronic offset correction should
be developed for stable operation under flight conditions.

\subsubsection{Background Spectrum}
\label{sec:czt2back}

In processing the CZT2 flight data it was immediately clear that
the recorded background spectra were much flatter than expected, failing to fall
off appreciably at high energy.  This was true also of the IMARAD pixels that
had been masked out due to their high noise: instead of just a low energy noise
peak, as expected, the masked pixels also had recorded a flat spectrum extending to
high energies.  This was probably due to cosmic ray events producing multiple triggers
in the ASIC, as we had no upper level discriminator implemented.  To produce 
accurate background spectra we therefore
averaged the masked pixels into a ``masked background'' spectrum and subtracted this from
the spectra of the active pixels.  The masked pixels were unfortunately all located 
on the IMARAD detector, so that it is not clear that this masked background is appropriate 
for the eV Products detector;
the resulting spectra from the two detectors agree reasonably well up to their
normalization, however 
(see below), so we do not believe this has had a big effect on our results.  In order to
compare these results with future data from the CZT3 detector, which will employ an ASIC
which provides only the peak channel for each event (see separate paper in these
proceedings), we have not added in charge 
from neighboring pixels in producing the CZT2 background spectra.

\begin{figure}
\begin{center}
\begin{tabular}{c}
\psfig{figure=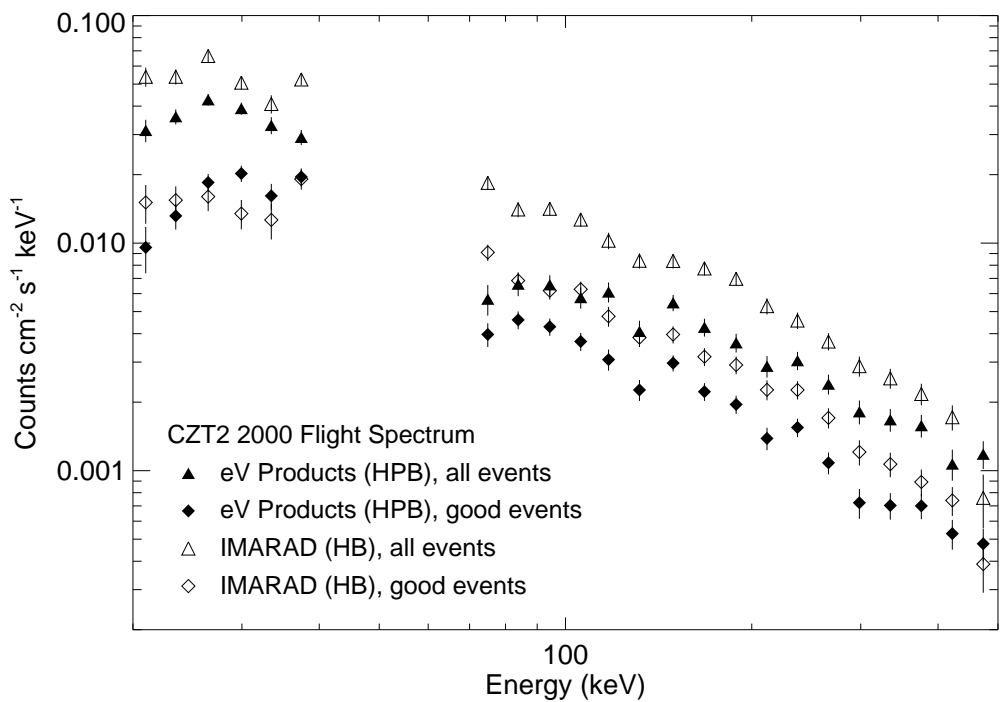,height=10cm}
\end{tabular}
\end{center}
\caption[initial]
{\label{fig:czt2data}
Background spectra recorded by the CZT2 experiment during the September 2000 flight.
Shown are both total and good events for the eV Products (HPB CZT) detector (closed
symbols) and the IMARAD (HB CZT) detector (open symbols).  The background level at 
100 keV is $\sim 4 \times 10^{-3}$ \cts ~for the eV Products detector and
$\sim 6 \times 10^{-3}$ \cts ~for the IMARAD detector.  In both cases the plastic
shield vetoes reduce the background by a factor of $\sim$ 2--2.5.  Events between
42 keV and 67 keV have been excluded due to poor subtraction of the 60 keV
$^{241}$Am calibration line.  
}
\end{figure}

The September 2000 flight spectra recorded by the CZT2 experiment are shown 
in Figure~\ref{fig:czt2data}.
Plotted are total and good (non-vetoed by the plastic shields) events for both the eV
Products HPB detector and the IMARAD HB detector.  A total of 4.9 hours of data
is included, during which time the temperature increased from 10 $^{\circ}$C to
17 $^{\circ}$C.  From Figure~\ref{fig:temp} we expect the offset to vary by 
less than 2 keV and the gain by less than 0.5 keV over this range.  Due to the coarse
binning used for the background continuum spectra we have not made these small 
corrections; rather, 
the individual spectra were simply shifted in energy by 4--5 keV so that the  
$^{241}$Am calibration lines were superimposed at 60 keV.  
The spectra represent the average of all pixels in
each detector with similar response and material quality (10 pixels from the eV Products
detector and 3 pixels from the IMARAD detector), as judged by the 
$^{241}$Am calibration line.  Due to the slight line broadening in flight noted 
in Section~\ref{sec:czt2performance}
it was difficult to cleanly subtract the calibration line profile, recorded on the ground,
from the flight background spectra.
We have therefore excluded events between 42 keV and 67 keV from the spectra shown.

The CZT2 spectra presented in Figure~\ref{fig:czt2data} show that simple passive and plastic
shielding can reduce background in CZT detectors to reasonable levels.  The
background recorded at 100 keV is $\sim 4 \times 10^{-3}$ \cts ~for the eV Products 
detector and
$\sim 6 \times 10^{-3}$ \cts ~for the IMARAD detector.  The spectra overall agree in shape
and normalization to within a factor of $\sim 2$.  The difference in normalization
could well arise from using the IMARAD masked pixel background for the eV Products 
crystal.  In both cases the plastic
shield vetoes reduce the background by a factor of $\sim$ 2--2.5.  Also notable is
the lack of any detected background lines, which a pixellated detector should be
able to resolve.  This is in agreement with the findings of the 
WUSTL/UCSD experiment\cite{slavis98}, which used crossed strips to achieve
energy resolution similar to that of a pixel detector. These findings cast doubt on the 
importance of activation lines from neutron interactions as a source
of CZT background\cite{parsons96,harrison98}, at least for lines below $\sim 500$ keV.

\subsection{Comparison of CZT1 and CZT2 Results}
\label{sec:compareczts}

\begin{figure}
\begin{center}
\begin{tabular}{c}
\psfig{figure=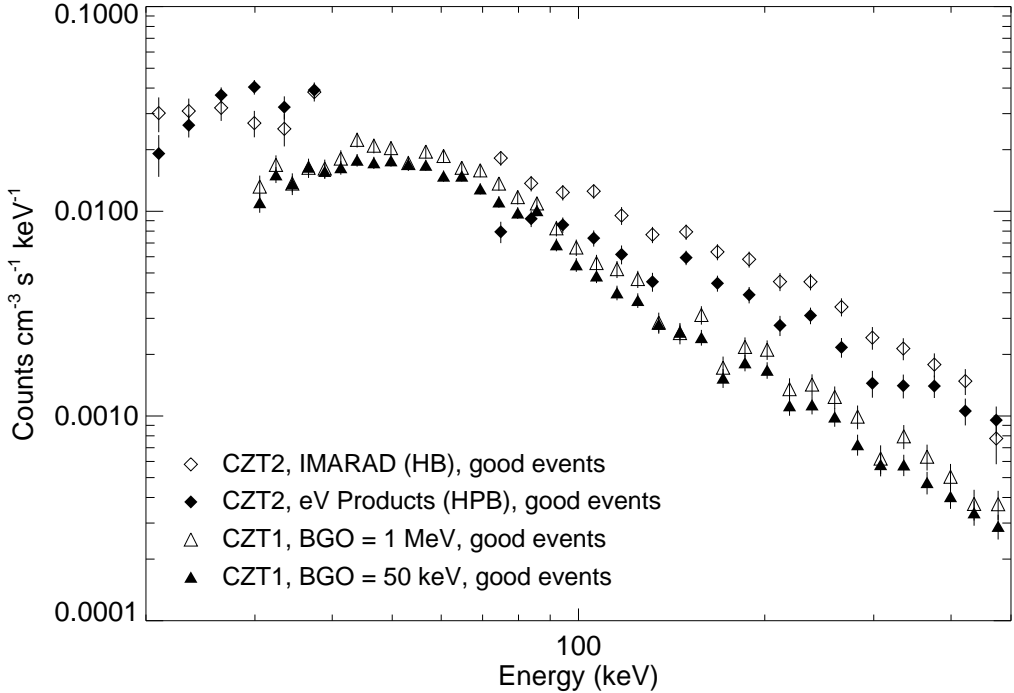,height=10cm}
\end{tabular}
\end{center}
\caption[initial]
{\label{fig:vol}
Comparison of CZT1 and CZT2 results.  Here the background per unit volume is presented
due to the different detector thicknesses.  Both CZT2 detectors are shown, as is the
CZT1 rate for both BGO threshold settings.  At high energies the CZT1 background is up to
a factor of $\sim 4$ lower than the CZT2 background, presumably due to the large BGO active
shield.  This difference is relatively modest, however, and the far lower mass of 
plastic shielding may make this a worthwhile tradeoff for many applications.
}
\end{figure}

In Figure~\ref{fig:vol} we compare the flight spectra recorded by CZT1 and CZT2,
plotted as count rate per volume to account for the different detector thicknesses.
Both CZT2 detectors are plotted, as are CZT1 results for both BGO threshold settings.
The agreement below 100 keV is quite good.  At higher energies the CZT2 spectra are
flatter, reaching
up to a factor of $\sim 4$ higher than the CZT1 spectra.  This is not surprising; 
clearly the large BGO crystal is more effective at shielding the small CZT1 detector
than plastic, even at a threshold of 1 MeV.  CZT2 also unavoidably had passive material
in its field of view,
which might contribute to the difference as well.
However, the relatively modest increase 
in background for CZT2 with passive/plastic shielding only may be a small price to
pay compared to the savings in space and mass achieved with thin, light plastic
scintillator shields, especially for large, wide-angle, coded-mask survey telescopes.
These results again support the findings of the WUSTL/UCSD experiment\cite{slavis99}.

\section{Detector Simulations and Predicted Background}
\label{sec:sim}

We have attempted to reproduce the results of both CZT experiments using Monte Carlo 
simulations conducted with MGEANT\cite{mgeant}, an improved user interface to the 
standard CERN Program Library simulation package GEANT.  Mass models of both CZT1 and
CZT2 were constructed and illuminated with parameterizations of the atmospheric
gamma-ray spectrum\cite{gehrels}, cosmic ray proton spectrum\cite{swartz}, and 
atmospheric neutron spectrum\cite{armstrong73,swartz}.  These input spectra were
assumed to be isotropic for simplicity, and are generally accepted to be accurate 
to about a factor of 2.  The location and energy deposit
of all interactions in the CZT and shields were recorded.  The CZT energy deposits
were then modified to take into account the response of the detectors.  For CZT1, a 
planar detector, this was done simply according to the Hecht relation\cite{hecht32} using
charge transport values of $\mu_e\tau_e = 5 \times 10^{-3}$
cm$^2$ V$^{-1}$ and $\mu_h\tau_h = 3 \times 10^{-5}$cm$^2$ V$^{-1}$\cite{bloser98}.
For CZT2 the deposited charge was propagated according to the applied electric field
and the induced signal calculated from the weighting potential; we used
$\mu_e\tau_e = 3 \times 10^{-3}$ cm$^2$ V$^{-1}$ and $\mu_h\tau_h = 10^{-6}$cm$^2$ 
V$^{-1}$\cite{bloser2000}.  These $\mu\tau$ values were derived for each detector
by matching simulated line spectra to recorded line spectra.
The resulting spectra were convolved with the measured
energy resolution of each detector.

\begin{figure}
\begin{center}
\begin{tabular}{c}
\psfig{figure=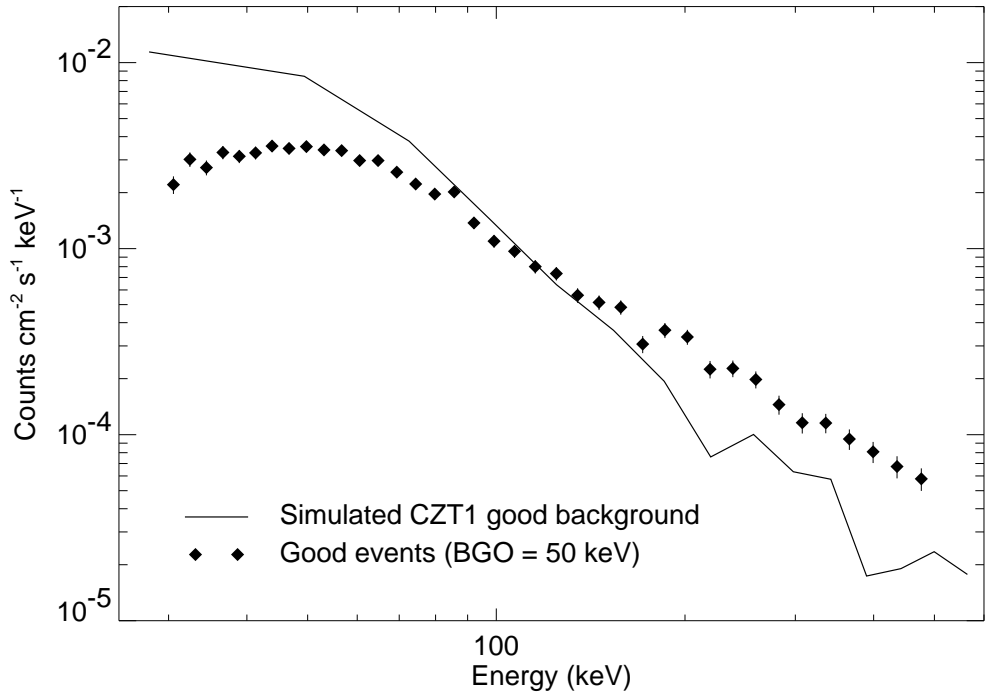,height=10cm}
\end{tabular}
\end{center}
\caption[initial]
{\label{fig:czt1sim}
CZT1 Good event flight spectrum (BGO threshold = 50 keV) compared to MGEANT simulation.
Only gamma-ray-induced events are included, as the simulated contribution from protons 
was completely vetoed by the plastic and BGO shields.  At high energies, the simulation 
fails to reproduce the recorded counts, much the same as in the 1997 
flight\cite{bloser98}.  At low energies the predicted level of aperture flux is not seen,
perhaps due to shielding of the CZT by the gondola structure (not included in the 
simulations). 
}
\end{figure}

The good event flight spectrum of CZT1 (with a BGO threshold of 50 keV) is compared to 
the MGEANT simulation in Figure~\ref{fig:czt1sim}.
In the simulation all proton-induced events were vetoed by the plastic and BGO and 
the neutron component was
negligible (see below), so only the gamma-ray component is shown.  Although the 
simulations agree with the data near 100 keV, it is clear that the overall shape is not
well-reproduced.  At high energies the simulations fail to reproduce the recorded
counts by a factor of 2--3, much the same as in the 1997 flight\cite{bloser98}.  Given
the large uncertainties in the input atmospheric gamma-ray spectrum, this is not too
surprising.  At low energies the predicted aperture flux, enhanced by the Hecht relation, 
is not observed, perhaps
due to shielding by the gondola structure (which was not included in the simulation
due to its size and complexity).  The CZT1 instrument was bolted to the main EXITE2
elevation flange next to an electronics rack, both sizable pieces of aluminum, and 
it is possible that cosmic ray or neutron-induced background in this structure 
contribute to the higher-than-predicted count rate at high energies.  

\begin{figure}
\begin{center}
\begin{tabular}{c}
\psfig{figure=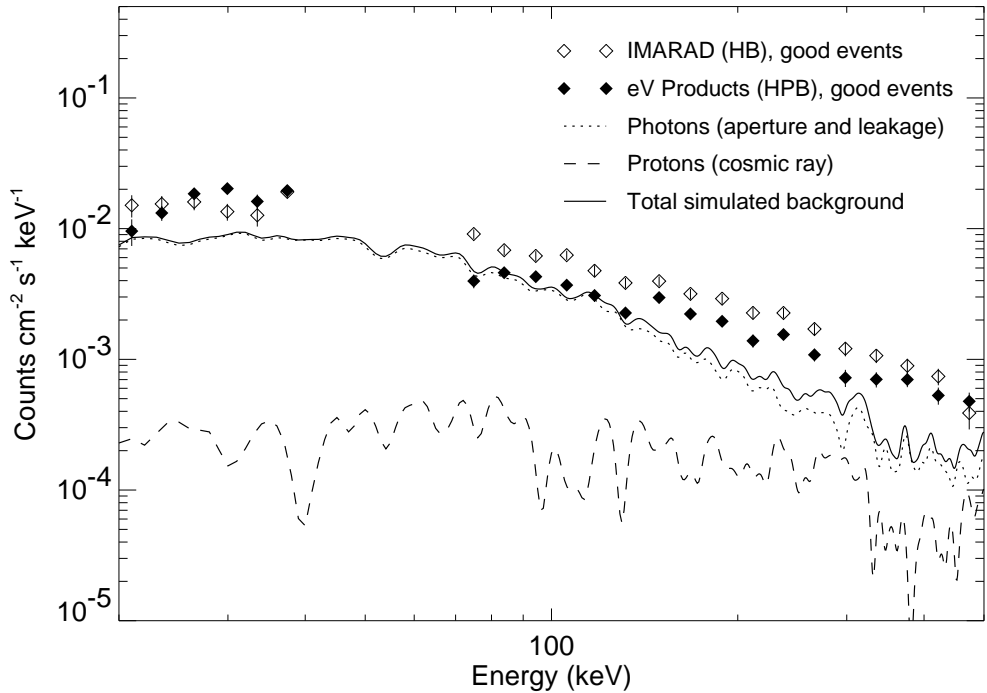,height=10cm}
\end{tabular}
\end{center}
\caption[initial]
{\label{fig:czt2sim}
CZT2 flight spectra (eV Products and IMARAD detectors) compared to the MGEANT 
simulations.  Included are the predicted spectra from aperture photons, shield
leakage photons, and cosmic ray proton interactions.  The contribution from
atmospheric neutrons was found to be negligible, although the standard 
MGEANT-GCALOR setup is probably too simplistic to be realistic.  
The jagged structure in the proton spectrum is due to poor statistics 
convolved with the detector energy resolution.
The data and
simulation agree to within a factor of $\sim 2$, although the recorded spectrum
is systematically flatter at high energies, as in CZT1.
}
\end{figure}

In Figure~\ref{fig:czt2sim} we compare the measured CZT2 flight spectra to the
MGEANT simulation.
We show separately the predicted background generated by atmospheric gamma-rays,
both via aperture flux and shield leakage, and the background from cosmic ray
proton interactions.  The contributions from atmospheric neutrons were again 
negligible.  We note that accurate simulations of neutron-generated background are
notoriously difficult to perform\cite{harrison98}, and that the standard 
MGEANT-GCALOR setup we used is probably too simplistic to be believable without
very careful attention to the precise isotopic composition of the detectors and 
surrounding materials.  The combined photon and proton simulated spectra agree with
the data to within a factor of $\sim 2$, falling systematically shy of the 
mark at high energies as in CZT1.  The CZT2 experiment was surrounded by extensive
gondola and detector structures, and background generated in these could play a 
role in the flatter observed spectrum.  We did take care to include two thick pieces of
aluminum in the CZT2 field of view.  We note that, although no lines are seen below
500 keV, higher-energy neutron activation lines cannot be excluded, and these would
form a Compton continuum in the CZT which might account for the higher-than-predicted
high energy background in both CZT1 and CZT2.  
Considering the uncertainty in the input spectra,
however, we consider the general level of agreement between the data and simulations
to be satisfactory.

\section{Conclusions and Future Work}
\label{sec:conc}

We have successfully flown two CZT detector experiments, dubbed CZT1 and CZT2,
that have provided valuable information about detector technology, shielding, and
background mechanisms.  CZT2 was the first array of pixellated CZT detectors, read
out by an ASIC, to be flown on a balloon, and our results have demonstrated that 
the basic technical procedures envisioned for the EXIST concept are sound.  The
CZT1 data, while again proving the necessity of active shielding, provide further 
evidence that this shielding need not be of the heavy, inorganic scintillator type,
but may rather be lighter plastic shielding to veto charged particles.  This argues
against prompt neutron reactions being a major source of background in CZT.  
The CZT2 spectra revealed no gamma-ray lines below 500 keV from activated isotopes, 
and the basic agreement with the CZT1 data and with simulations shows again
that a combination of passive and plastic shielding can provide reasonable background
levels.  Our work on advanced CZT detectors for astronomical applications will continue
with the development of CZT3, employing a $2 \times 2$ array of larger detectors with
a more advanced contacting and readout scheme, as described separately in these
proceedings.

\acknowledgments     
 
We would like to thank the entire NSBF team for a successful balloon
flight, J. Apple and K. Dietz for technical support in the field, and B. Sundal 
for help with the VA-DAQ system. 
This work was supported in 
part by NASA grants NAG5-5103 and NAG5-5209.  


\bibliography{spie2001_paper}   
\bibliographystyle{spiebib}   
 
\end{document}